\documentclass[namedreferences]{solarphysics}

\usepackage[linksfromyear,optionalrh]{spr-sola-addons} % For Solar Physics 
\usepackage{graphicx}                    % For eps figures, newer & more powerfull
\usepackage{type1cm}
\usepackage{lmodern}
\usepackage{multirow}
\usepackage{url}
\usepackage{amsmath} 
\newcommand{\aap}{    {\it Astron. Astrophys.}}

\newcommand{\apj}{    {\it Astrophys. J.}}

\newcommand{\solphys}{{\it Solar Phys.}}

\begin{document}
\begin{article}

%----------------------------------------------------------------------------------------------------
\begin{opening}

\title{Oscillations and Waves in Radio Source of Drifting Pulsation Structures}

\author[ addressref={aff1}]{\inits{M.}\fnm{Marian}~\lnm{Karlick\'{y}}\orcid{0000-0002-3963-8701}}
\author[ addressref={aff2}, corref, email={rybak@astro.sk}]{\inits{J.}\fnm{J\'{a}n}~\lnm{Ryb\'{a}k}\orcid{0000-0003-3128-8396}}
\author[ addressref={aff1} ]{\inits{C.}\fnm{Miroslav}~\lnm{B\'arta}}

\runningauthor{M. Karlick\'y, J. Ryb\'ak, and M. B\'arta}
\runningtitle{Oscillations and Waves in Radio Source of DPSs}

\address[id=aff1]{Astronomical Institute, Academy of Sciences of
                  the Czech Republic, 251 65 Ond\v{r}ejov, Czech Republic}
\address[id=aff2]{Astronomical Institute, Slovak Academy of Sciences,
                  Tatransk\'{a} Lomnica, Slovakia}

%----------------------------------------------------------------------------------------------------
\begin{abstract}
Drifting pulsation structures (DPSs) are considered to be radio signatures of
the plasmoids formed during magnetic reconnection in the impulsive phase of
solar flares. In the present paper we analyze oscillations and waves
in seven examples of drifting pulsation structures, observed by the
800--2000\,MHz Ond\v{r}ejov radiospectrograph. For their analysis we use a new
type of oscillation maps which give us a much more information about
processes in DPSs than that in previous analyzes. Based on these oscillation
maps, made from radio spectra by the wavelet technique, we recognized
quasi-periodic oscillations with periods ranging from about 1 to 108\,s
in all studied DPSs. This strongly supports the idea that DPSs are generated
during a fragmented magnetic reconnection. Phases of most the oscillations in
DPSs, especially for the period around 1\,s, are synchronized ("infinite"
frequency drift) in the whole frequency range of DPSs. For longer
periods in some DPSs we found that the phases of the oscillations drift with
the frequency drift in the interval from $-17$ to
$+287$\,MHz\,s$^{-1}$. We propose that these drifting phases can be caused (a)
by the fast or slow magnetosonic waves generated during the magnetic
reconnection and propagating through the plasmoid, (b) by a quasi-periodic
structure in the plasma inflowing to the reconnection forming a plasmoid, and
(c) by a quasi-periodically varying reconnection rate in the X-point of the
reconnection close to the plasmoid.
\end{abstract}

\end{opening}

%----------------------------------------------------------------------------------------------------
\section{Introduction}

Drifting pulsation structure was for the first time recognized in the paper
by \cite{2000A&A...360..715K}. It was interpreted as radio emission from
the plasmoid formed during magnetic reconnection in the impulsive phase of
solar flares. Details of this interpretation were presented in the papers by
\cite{2007A&A...464..735K,2008A&A...477..649B,2008SoPh..253..173B,2010A&A...514A..28K,2011ApJ...737...24B,2011ApJ...733..107K}
and can be summarized as follows. In the flare current sheet, formed below the
rising magnetic rope during magnetic reconnection, plasmoids are generated
due to the tearing-mode/plasmoid  instability \citep{Loureiro+:2007}.
At the X-points of the magnetic reconnection, superthermal  electrons  are  
accelerated
\citep{2008ApJ...674.1211K}.  Then these electrons  are trapped  in  a  nearby
plasmoid,  where they  generate plasma waves. These waves are then converted to
the electromagnetic waves observed as  DPSs \citep{2010A&A...514A..28K}. The
narrow bandwidth of DPSs is caused by the limited interval of the plasma
densities (plasma frequencies) inside the plasmoid. Due to divergence of the
magnetic field lines in the upward direction in the solar atmosphere plasmoids
preferentially move upwards \citep{2008A&A...477..649B}, that is why DPSs
mostly drift toward low  frequencies.  The  velocity  of  plasmoids can be as high 
as the local Alfv\'en speed. The acceleration of electrons at X-points of magnetic
reconnection is believed to be quasi-periodic, which causes quasi-periodic
pulsations of DPSs.

As mentioned above, the plasmoid is an important part of the magnetic
reconnection process. Thus, the corresponding DPSs can be used for diagnostics of
processes in magnetic reconnection in solar flares. Some analysis of DPSs
were made in the papers by~\cite{2005A&A...432..705K, 2015SoPh..290..169D}. For
example, DPS observed during the 12 April 2001 flare with the high time
resolution (1\,ms) on the single frequencies of 610 and 1420\,MHz was analyzed
by the Fourier method and periods in the interval of 0.9--7.5\,s were
found~\citep{2005A&A...432..705K}. For an interval of periods shorter than 1\,s
the authors found that the periods are distributed according to the power-law
function with the power-law index from $-1.3$ to $-1.6$. The power-law distribution
with the similar power-law index was also found in the 0.025--0.1\,s period
interval in DPS from the 12 July 2000 flare, observed with the very high time
resolution (80\,$\mu$s) by the 1352--1490\,MHz \textit{Torun Radiospectrograph}
~\citep{2015SoPh..290..169D}.

In the present paper, we continue in the analysis of DPSs and thus processes in
the radio source of DPSs, {\it i.e.}, processes in plasmoids. But, compared
with the previous studies we use a much more sophisticated method, which was
for the first time used in the paper by~\cite{2017SoPh..292....1K}. This new
method gives us a more detailed information about waves and
oscillations in DPSs. We analyze a set of DPSs in order to obtain some
statistics about these processes.

The paper is organized as follows: In Section 2 we show data and describe
methods of their analysis. The results of the analysis are summarized in
Section~3. In Section 4 the results are interpreted and conclusions are
summarized in Section 5.

%----------------------------------------------------------------------------------------
\section{Data and Methods of Their Analysis}

\begin{table}[t!]
\caption{List of the studied DPSs and the parameters of the soft X-ray emission
of the related solar flares.}
\label{tab1} %\centering
\begin{tabular}{cccccc}
\hline\hline
DPS & Date         &  Flare start  & Flare maximum & Flare end  & GOES  \\
no. &              &  (UT)         &    (UT)       &  (UT)      & classification \\
\hline
1   & 1992 Oct.  5 & 09:13         & 09:31         & 09:56      & M2.0  \\
2   & 1998 Aug. 18 & 08:14         & 08:24         & 08:32      & X2.8  \\
3   & 2002 May  20 & 15:21         & 15:27         & 15:31      & X2.1  \\
4   & 2002 Aug. 30 & 12:47         & 13:29         & 13:35      & X1.5  \\
5   & 2003 May  27 & 05:06         & 06:26         & 07:16      & M1.6  \\
6   & 2005 Jun. 14 & 06:54         & 07:30         & 07:56      & C4.2  \\
7   & 2014 Apr. 18 & 12:31         & 13:03         & 13:20      & M7.3  \\
\hline
\end{tabular}
\end{table}

% instrument
The radio spectra of seven DPSs selected for our analysis, listed in
Table~\ref{tab1}, were acquired by \href{http://www.asu.cas.cz/~radio/}{\it
Ond\v{r}ejov Radiospectrograph} RT5 located at Ond\v{r}ejov, Czech Republic.
Radiospectrograph frequency range, frequency resolution, and temporal
resolution of the instrument are 800--2000\,MHz, 4.7\,MHz, and 10\,ms,
respectively \citep{2008SoPh..253...95J}.
% essence of the data reduction method before the wavelet analysis
No special data reduction has been applied to the archive data of the
RT5 radiospectrograph.
% Resampling
Data were only re-sampled to the final temporal sampling of 0.1\,s
(1\,s for DPS No.7).
% Bad frequencies
At some frequencies, man-made radio interference affected the acquired signal.
Such parts of the data are excluded and they are marked by black bands
in all plots of the radiospectrograms.
% plots
Plots of three radiospectrograms can be seen as the second panels in
Figures~\ref{figure1} (DPS No.1), \ref{figure3} (DPS No.3), and \ref{figure5}
(DPS No.7).

% method
For analysis of the radio spectra we utilize a novel method introduced in
articles of \cite{2017SoPh..292....1K} and \cite{2017SoPh..292...94K}. This
method is based on the wavelet transform (WT) \citep{1998BAMS...79...61T}
providing a clear detection of time--frequency evolution of the strong radio
signal wave-patterns. The statistically significant wave-pattern signal power
is identified in the selected period range at any temporal moment and radio
frequency. Then the method provides an overplot of the phases of the detected
significant radio signal wave-patterns at their temporal and frequency
locations on a plot of the original radiospectrogram. Due to possible
multiplicity of the signal periodicities the investigated periodicities have to
be limited to narrower WT period ranges of particular interest.

In this work the Morlet mother wavelet, consisting of a complex sine
wave modulated by a Gaussian, is selected to search for radio signal
variability, with the non-dimensional frequency $\omega_0$ satisfying
the admissibility condition~\citep{1992AnRFM..24..395F}.
The WT is calculated for the period range from 0.1 to 20\,s
(1 to 200\,s for DPS No.7) with scales sampled as fractional
powers of two with ${\delta j = 0.4}$, which is small enough to give
an adequate sampling in scale, a minimum scale ${s_{j}=0.096}$\,s
(0.96\,s for DPS No.7) and a number of scales $N=101$.
Both the calculated significance of the derived WT periodicities
and the cone-of-influence are taken into account as described in
\cite{2017SoPh..292....1K}.
The value of the confidence level was set to 99\,\%.

%----------------------------------------------------------------------------------------
\section{Results}

\begin{table}[t!]
\caption{Radio emission parameters of studied DPSs.}
\label{tab2} %\centering
\begin{tabular}{ccccc}
\hline\hline
DPS &Date          & Time  & Frequency & Frequency \\
no. &              &       &  range    & drift    \\
    &              & (UT)  &  [GHz]    & [MHz\,s$^{-1}$]\\
\hline
1   & 1992 Oct.  5 & 09:24:20--09:25:00 & 1.0--1.7 & $-12.1$ \\
2   & 1998 Aug. 18 & 08:18:20--08:18:40 & 0.8--1.4 & $-7.1$ \\
3   & 2002 May  20 & 15:24:55--15:25:28 & 1.0--1.7 & $-25.0$  \\
4   & 2002 Aug. 30 & 13:27:38--13:27:50 & 0.8--1.5 & $-30.0$  \\
5   & 2003 May  27 & 06:42:36--06:43:28 & 0.8--1.3 & $-6.3$  \\
6   & 2005 Jun. 14 & 07:21:32--07:22:20 & 1.0--1.5 & $-5.5$  \\
7   & 2014 Apr. 18 & 12:44:00--12:53:00 & 0.8--1.6 & $-1.66$  \\
\hline
\end{tabular}
\end{table}

\begin{table}[t!]
\caption{Main oscillation periods and parameters of the most distinct
drifting phases (DPHA) of oscillations found in DPSs.
The DPHAs are marked by the white arrows in Figures~\ref{figure2},
\ref{figure4} and \ref{figure6}.}
\label{tab3} \centering
\begin{tabular}{cccccc}
\hline\hline
DPS & Main oscillation & DPHA  & DPHA      & DPHA            & DPHA    \\
 No.& periods          & time  & frequency & freq. drift     & period  \\
    & [s]              & (UT)  & [GHz]     & [MHz\,s$^{-1}$]  & [s]      \\
\hline
1  & 0.9--1.7, 2.4,    & 9:24:27.2--9:24:28.0 & 1.47--1.7  & $+287$ & 6.5--8.5\\
   & 3.9, 4.8, 5.2,    & 9:24:43.5--9:24:44.7 &  1.1--1.25 & $+125$ & 6.5--8.5\\
   & 7.8, 9.2 \\
\hline
2 & 0.5--1.3, 2.0, & no \\
  & 3.0, 3.5, 6.9, & DPHA \\
  & 9.1, 13.3 \\
\hline
3 & 1.0--1.6, 2.2, & 15:25:11.2--15:25:12.4 & 1.35--1.5 & $-125$ & 7.2--10.5\\
  & 2.9, 3.2, 4.9 \\
  & 5.7, 8.9 \\
\hline
4 & 0.7, 1.0, 2.0, & no \\
  & 3.0            & DPHA \\
\hline
5 & 0.5--1.5, 1.7,2.8, & no \\
  & 3.8--5.6, 8.5      & DPHA\\
\hline
6 & 0.5--1.0, 1.1, 2.1, & no \\
  & 3.0, 4.1, 5.9      & DPHA \\
\hline
7 & 18, 30, 65, & 12:46:30--12:46:48 & 1.2--1.5  & $-17$ & 65--115\\
  & 90, 108     & 12:49:18--12:49:36 & 1.1--1.47 & $-21$ & 63--75\\
\hline
\end{tabular}
\end{table}

We analyzed the DPSs observed in seven solar flares, which parameters are shown in
Tables~\ref{tab1} and \ref{tab2}.
From the comparison of Table 1 and Table 2 it can be seen, except the 27 May 2003 event
(DPS No.5), that the DPSs were observed before the GOES soft X-ray flare maximum.

An overall inspection of the radio spectra shows that all these DPSs drift
towards low frequencies with the frequency drift in the range from
$-1.66$ to $-30$\,MHz\,s$^{-1}$ (Table~\ref{tab2}).

Using the method described in the previous section we constructed for all DPSs
oscillation maps of the phases for intervals around all detected
statistically significant main oscillation periods.
These  main oscillation periods are listed in Table~\ref{tab3} (second column).

\begin{figure}[t!]
\centering
% histograms
\includegraphics[width=12.0cm,height=4.0cm, clip=true]
%period_histogram_all_and_selected_freq_1000-2000_MHz_time_092410-092500_per_0_5-0_9_s_lowres.eps}
{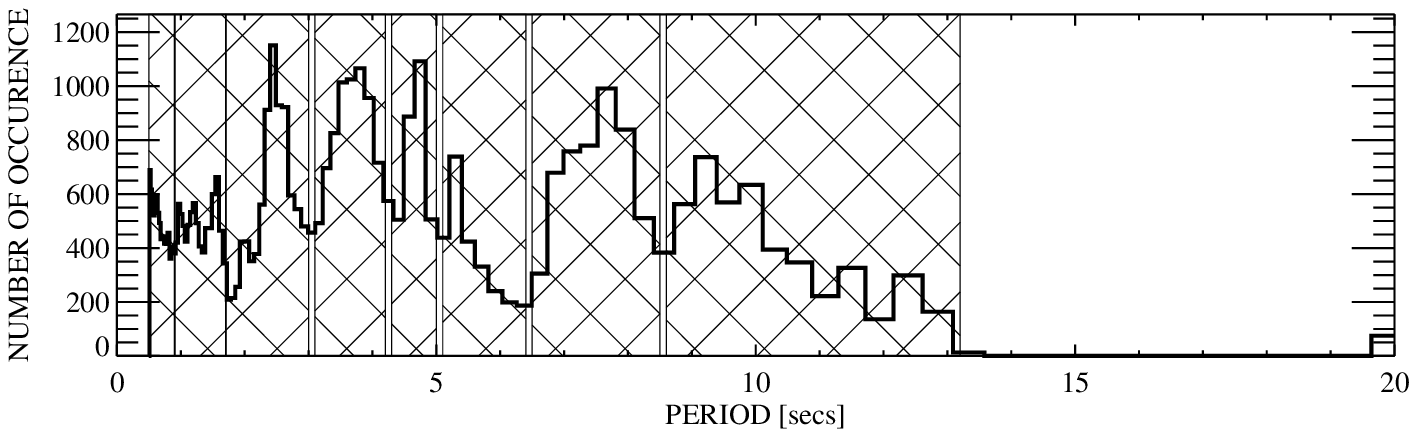}
% original spectrograms
\includegraphics[width=12.0cm,height=3.2cm, bb =30 30 400 135, clip=true]
%radio_base_WT_no_overplot_freq_1000-2000_MHz_time_092410-092500_lowres.eps}
{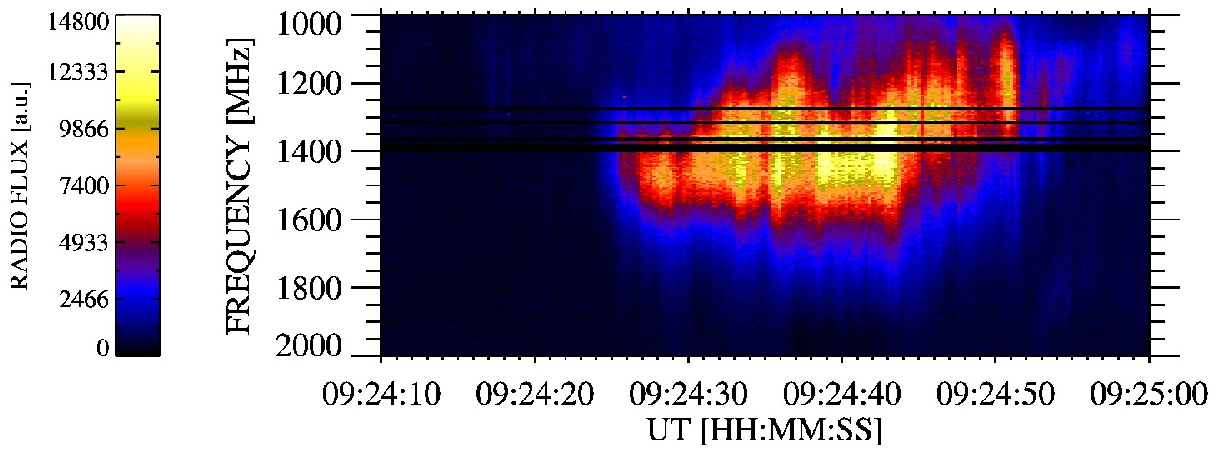}
% maps
\includegraphics[width=12.0cm,height=3.2cm, bb =30 30 400 135, clip=true]
%radio_base_WT_phase_overplot_freq_1000-2000_MHz_time_092410-092500_per_0_5-0_9_s_lowres.eps}
{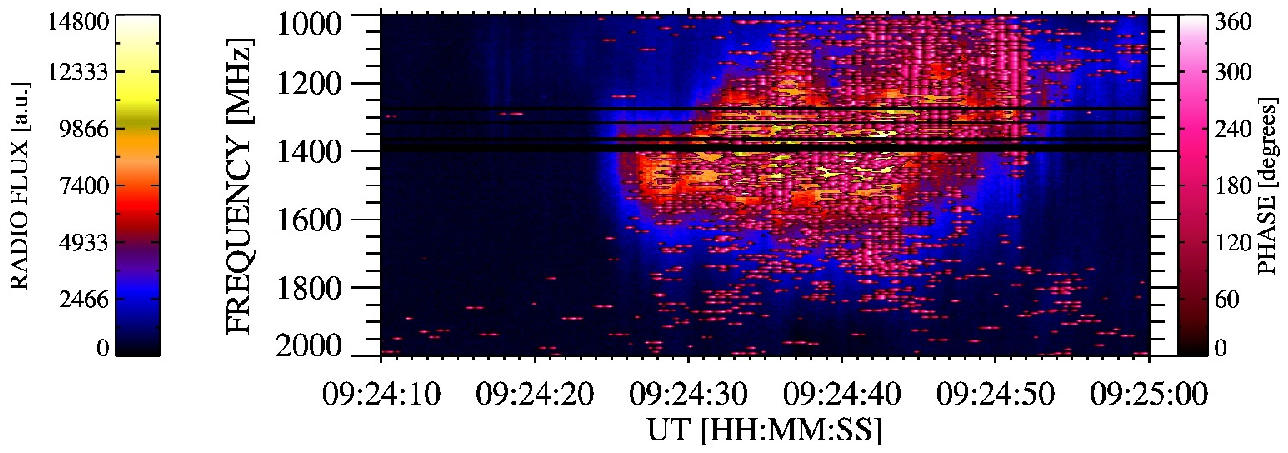}
\includegraphics[width=12.0cm,height=3.2cm, bb =30 30 400 135, clip=true]
%radio_base_WT_phase_overplot_freq_1000-2000_MHz_time_092410-092500_per_0_9-1_7_s_lowres.eps}
{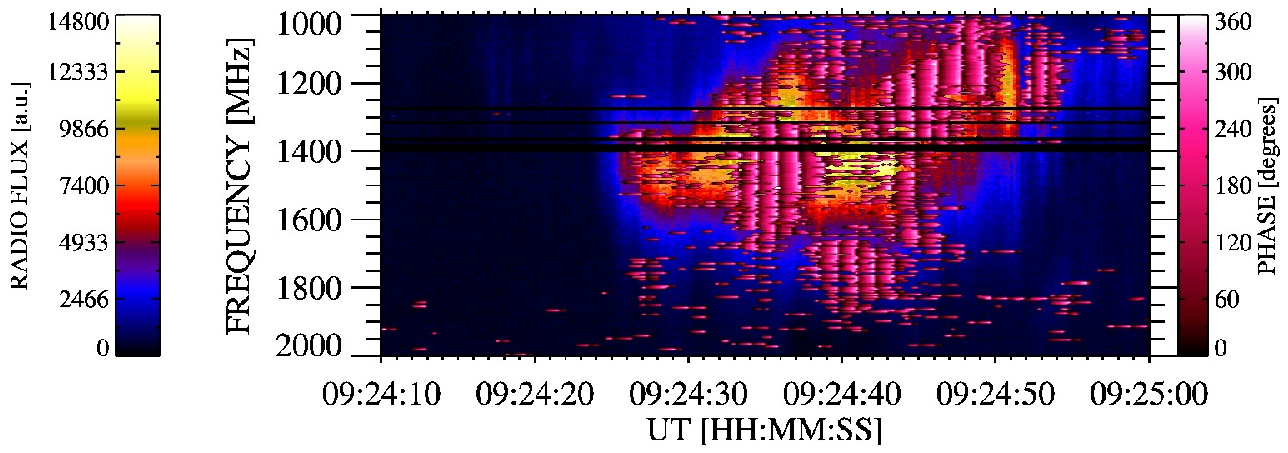}
\includegraphics[width=12.0cm,height=3.8cm, bb =30 10 400 135, clip=true]
%radio_base_WT_phase_overplot_freq_1000-2000_MHz_time_092410-092500_per_1_7-3_0_s_lowres.eps}
{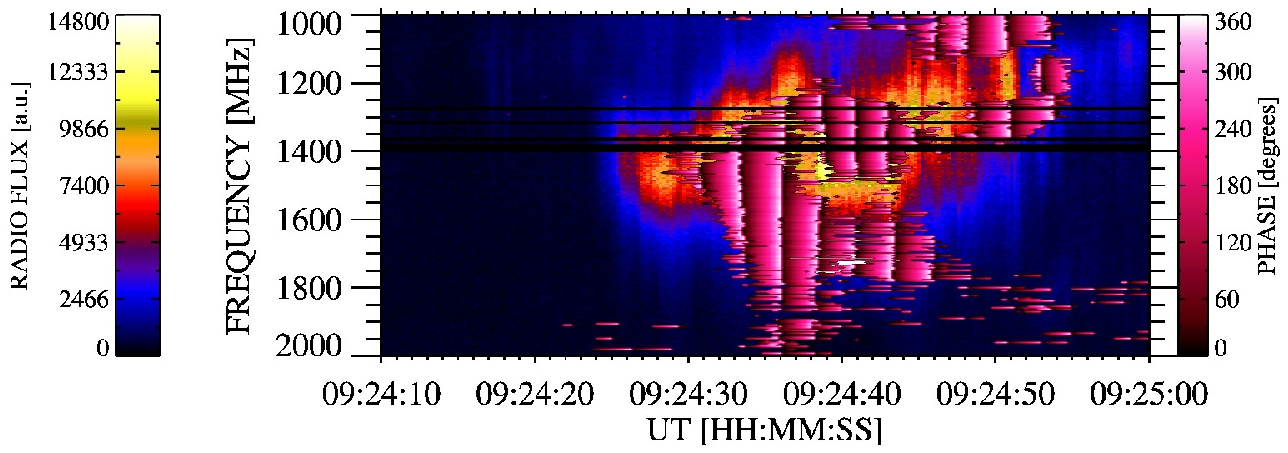}
\caption{DPS in the 05 October 1992 flare~--- time interval 09:24:10--09:25:00\,UT:
\textit{first panel} (\textit{from up to down}):
histogram of the WT significant periods, where the ranges of periods
selected for the following oscillation maps are expressed by hatched areas;
\textit{second panel}: the radio spectrum observed by
the \textit{Ond\v{r}ejov Radiospectrograph}; \textit{bottom three panels}: the
phase maps (\textit{pink areas with the black lines} showing the zero phase of 
oscillations) overplotted on the radio
spectrum for periods detected in the range of 0.5--0.9,
0.9--1.7, and 1.7--3.0\,s.}
\label{figure1}
\end{figure}

\begin{figure}[t!]
\centering
\includegraphics[width=12.0cm,height=3.2cm, bb =30 30 400 135, clip=true]
%radio_base_WT_phase_overplot_freq_1000-2000_MHz_time_092410-092500_per_3_1-4_2_s_lowres.eps}
{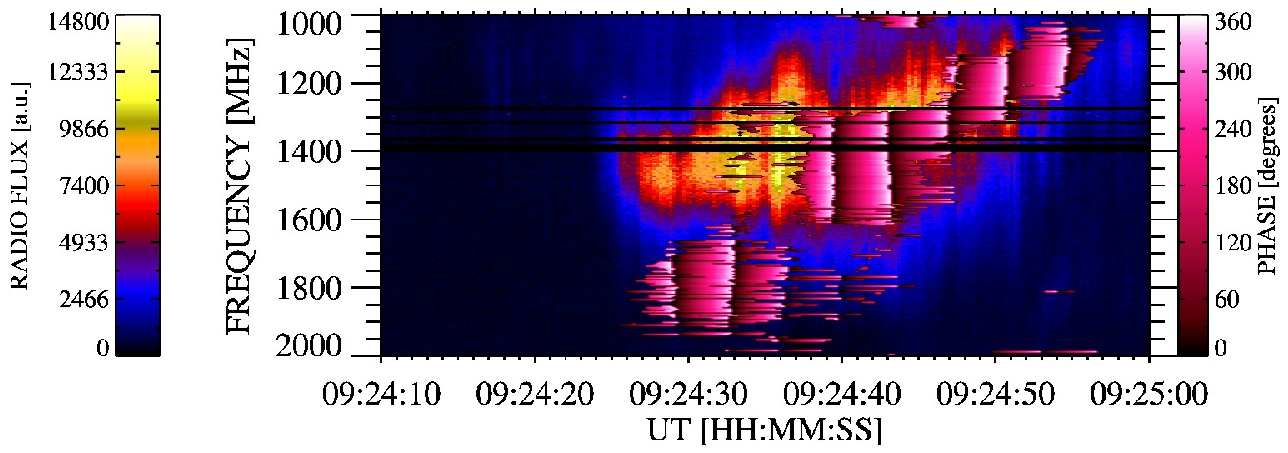}
\includegraphics[width=12.0cm,height=3.2cm, bb =30 30 400 135, clip=true]
%radio_base_WT_phase_overplot_freq_1000-2000_MHz_time_092410-092500_per_4_3-5_0_s_lowres.eps}
{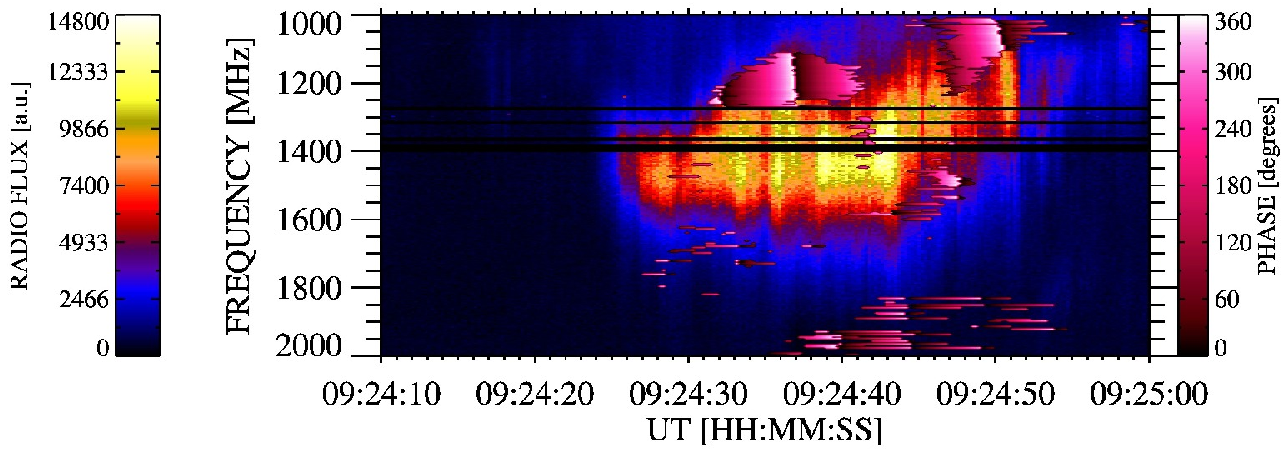}
\includegraphics[width=12.0cm,height=3.2cm, bb =30 30 400 135, clip=true]
%radio_base_WT_phase_overplot_freq_1000-2000_MHz_time_092410-092500_per_5_1-6_4_s_lowres.eps}
{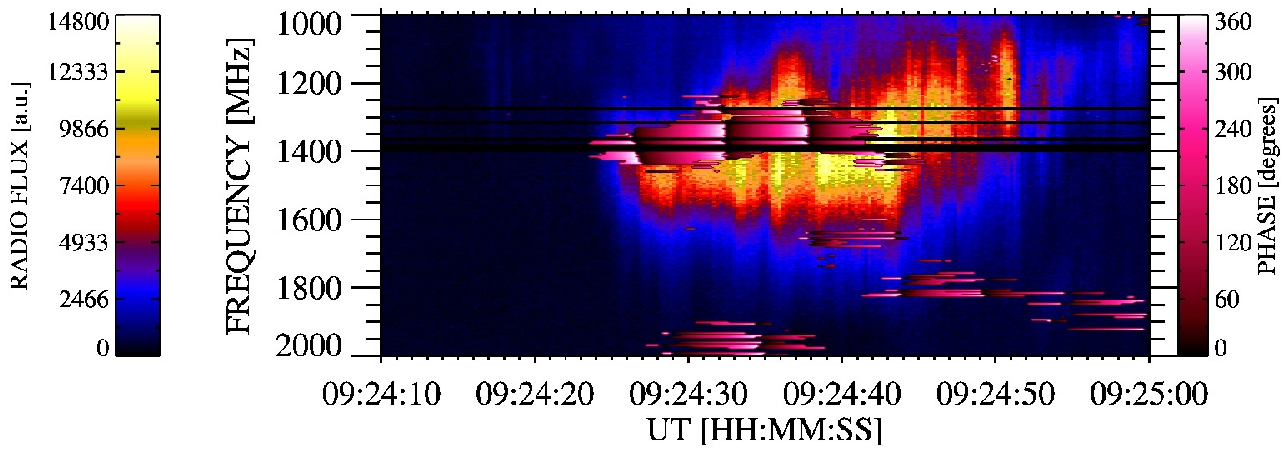}
\includegraphics[width=12.0cm,height=3.2cm, bb =30 30 400 135, clip=true]
%radio_base_WT_phase_overplot_freq_1000-2000_MHz_time_092410-092500_per_6_5-8_5_s_lowres_arrows.eps}
{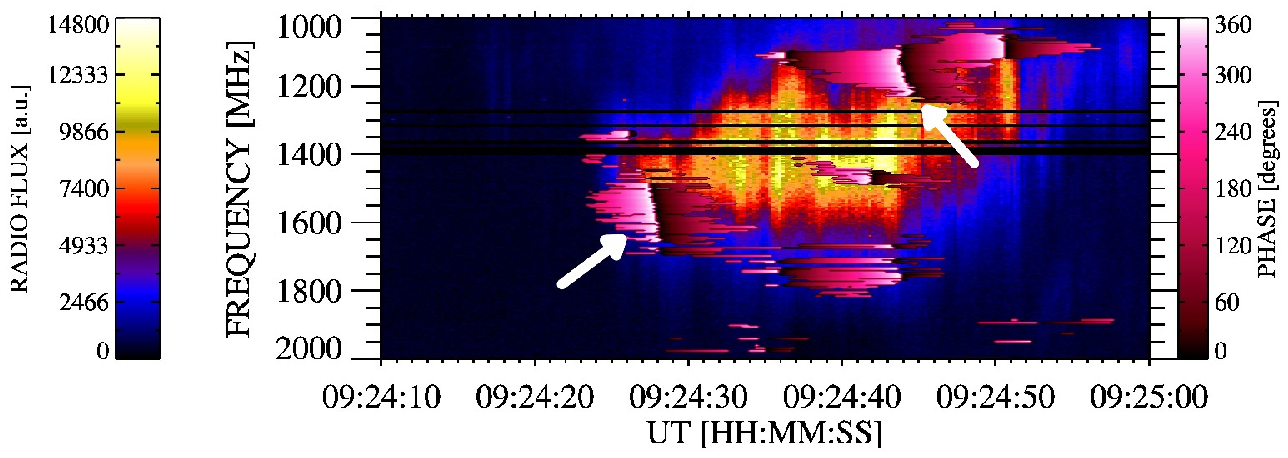}
\includegraphics[width=12.0cm,height=3.8cm, bb =30 10 400 135, clip=true]
%radio_base_WT_phase_overplot_freq_1000-2000_MHz_time_092410-092500_per_8_6-13_2_s_lowres.eps}
{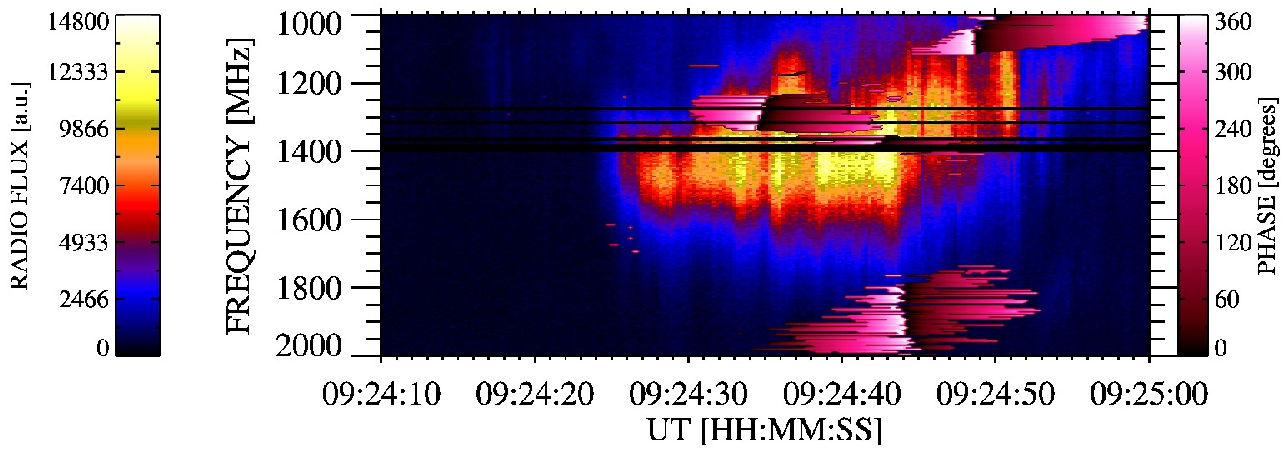}
\caption{DPS in the 05 October 1992 flare~--- continuation of Figure~\ref{figure1} for longer periods:
the phase maps (\textit{pink areas with the black lines} showing the zero phase of oscillations) overplotted
on the radio pectrum for periods detected in the range of 3.1--4.2, 4.3--5.0, 5.1--6.4,
6.5--8.5, and 8.6--13.2\,s (\textit{from up to down}).
The \textit{white arrows} show selected distinct drifting oscillation phases.
} \label{figure2}
\end{figure}

\begin{figure}[t!]
\centering
% histograms
\includegraphics[width=12.0cm,height=4.0cm, clip=true]
%period_histogram_all_and_selected_freq_800-2000_MHz_time_152430-152630_per_0_5-1_2_s_lowres.eps}
{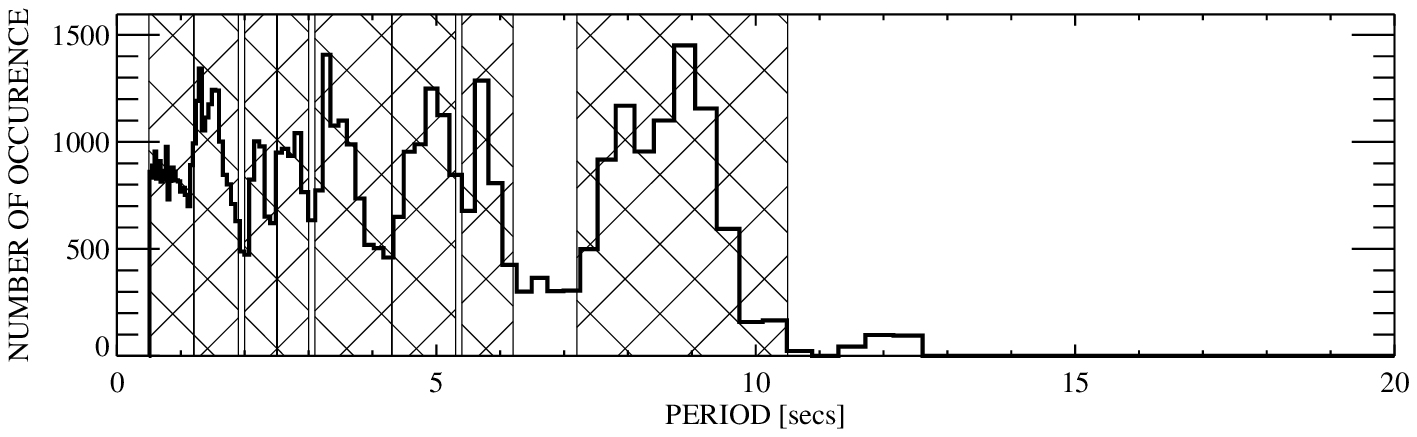}
% original spectrogram
\includegraphics[width=12.0cm,height=3.2cm, bb =30 30 400 135, clip=true]
%radio_base_WT_no_overplot_freq_800-2000_MHz_time_152430-152630_lowres.eps}
{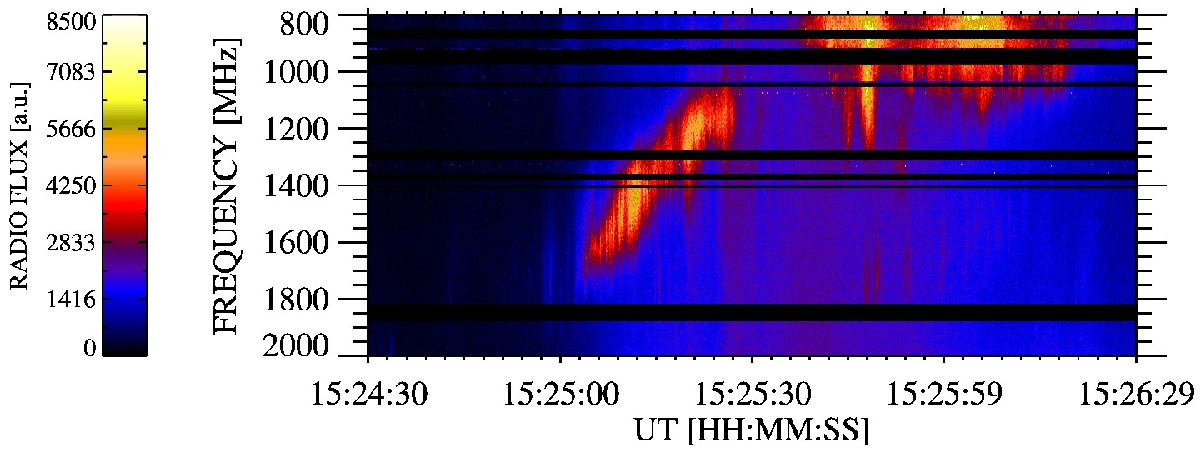}
% maps
\includegraphics[width=12.0cm,height=3.2cm, bb =30 30  400 135, clip=true]
%radio_base_WT_phase_overplot_freq_800-2000_MHz_time_152430-152630_per_0_5-1_2_s_lowres.eps}
{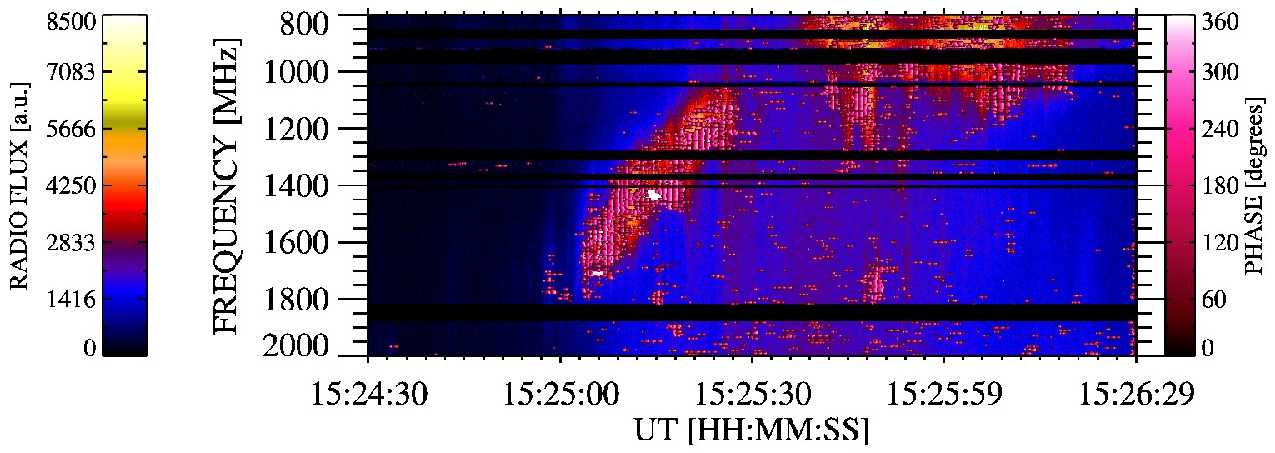}
\includegraphics[width=12.0cm,height=3.2cm, bb =30 30  400 135, clip=true]
%radio_base_WT_phase_overplot_freq_800-2000_MHz_time_152430-152630_per_1_2-1_9_s_lowres.eps}
{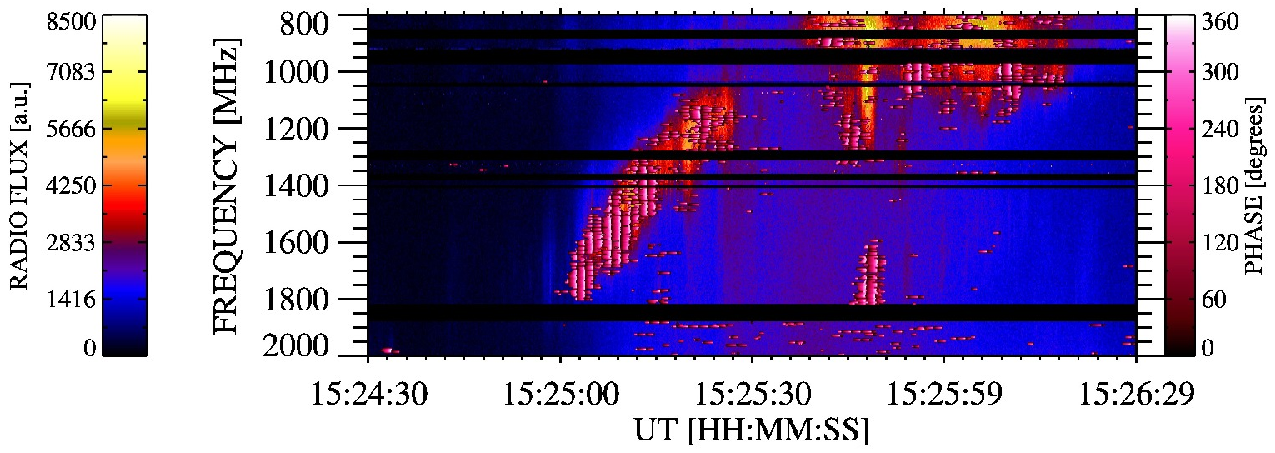}
\includegraphics[width=12.0cm,height=3.8cm, bb =30 10  400 135, clip=true]
%radio_base_WT_phase_overplot_freq_800-2000_MHz_time_152430-152630_per_2_0-2_5_s_lowres.eps}
{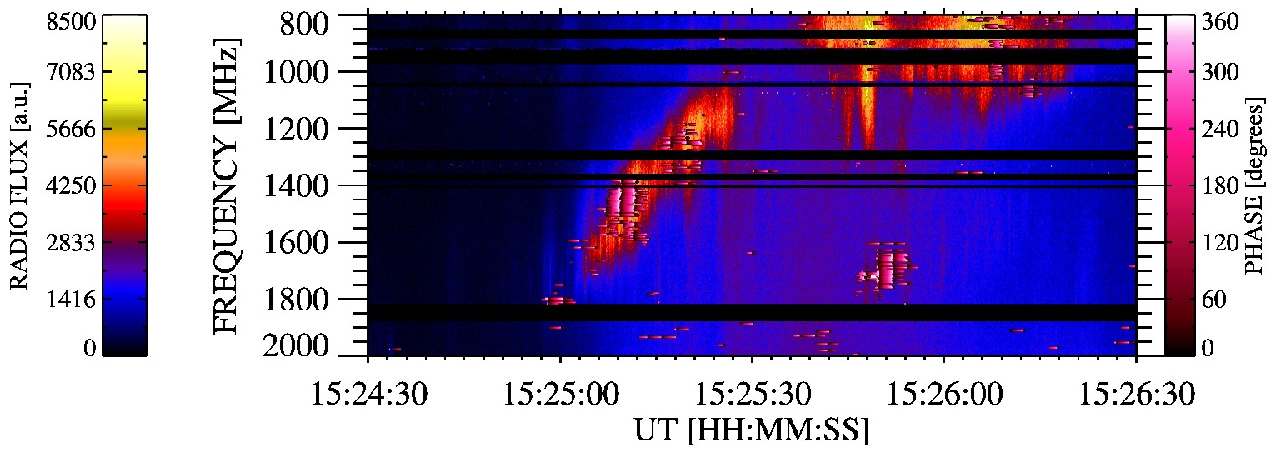}
\caption{DPS in the 20 May 2002 flare~--- time interval 15:24:30--15:26:30\,UT:
\textit{first panel} (\textit{from up to down}):
histogram of the WT significant periods, where the ranges of periods selected for the
following oscillation maps are expressed by \textit{hatched areas}; 
\textit{second panel}: the radio spectrum
observed by the \textit{Ond\v{r}ejov Radiospectrograph}; \textit{bottom three panels}:
the phase maps (\textit{pink areas with the black lines} showing the zero phase of 
oscillations) overplotted on
the radio spectrum for periods detected in the range of 0.5--1.2, 1.2--1.9, and
2.0--2.5\,s.} \label{figure3}
\end{figure}

\begin{figure}[t!]
\centering
\includegraphics[width=12.0cm,height=3.2cm, bb =30  30  400 135, clip=true]
%radio_base_WT_phase_overplot_freq_800-2000_MHz_time_152430-152630_per_2_5-3_0_s_lowres.eps}
{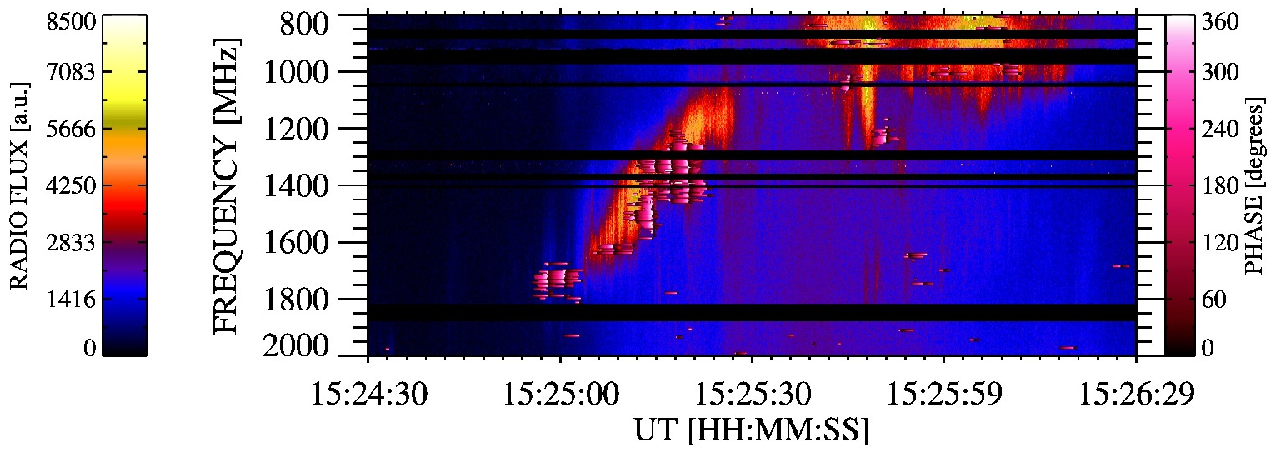}
\includegraphics[width=12.0cm,height=3.2cm, bb =30  30  400 135, clip=true]
%radio_base_WT_phase_overplot_freq_800-2000_MHz_time_152430-152630_per_3_1-4_3_s_lowres.eps}
{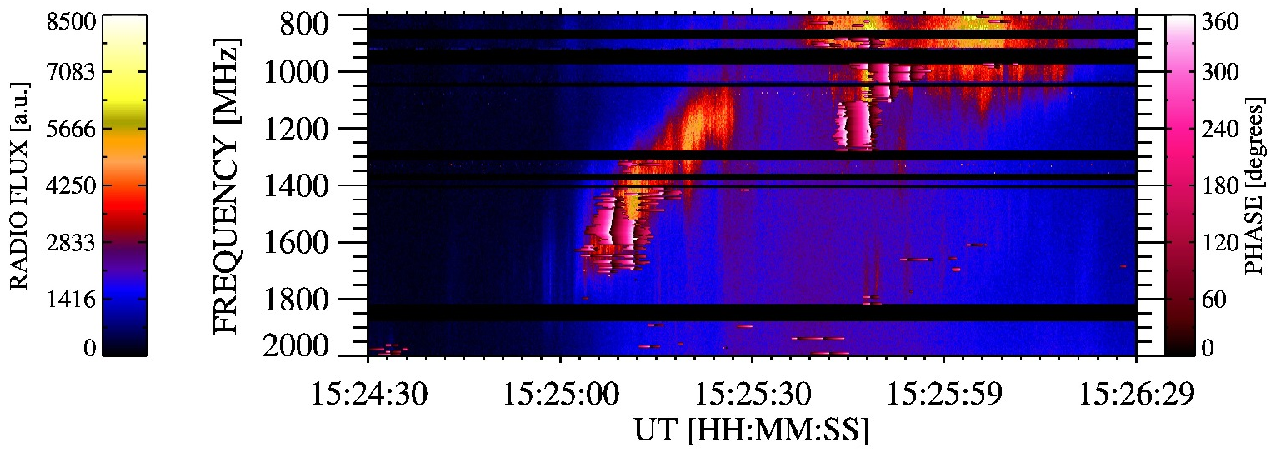}
\includegraphics[width=12.0cm,height=3.2cm, bb =30  30  400 135, clip=true]
%radio_base_WT_phase_overplot_freq_800-2000_MHz_time_152430-152630_per_4_3-5_3_s_lowres.eps}
{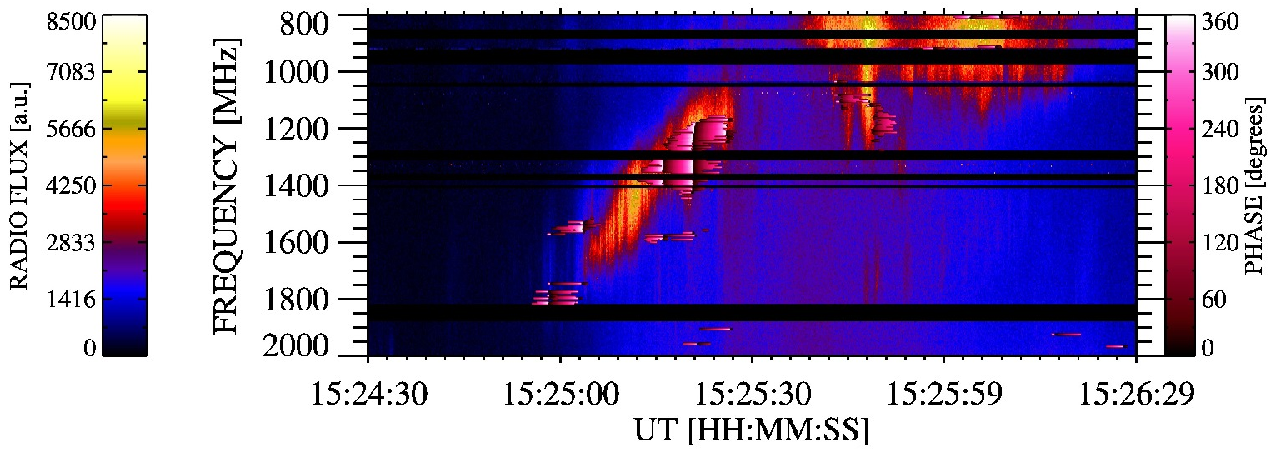}
\includegraphics[width=12.0cm,height=3.2cm, bb =30  30  400 135, clip=true]
%radio_base_WT_phase_overplot_freq_800-2000_MHz_time_152430-152630_per_5_4-6_2_s_lowres.eps}
{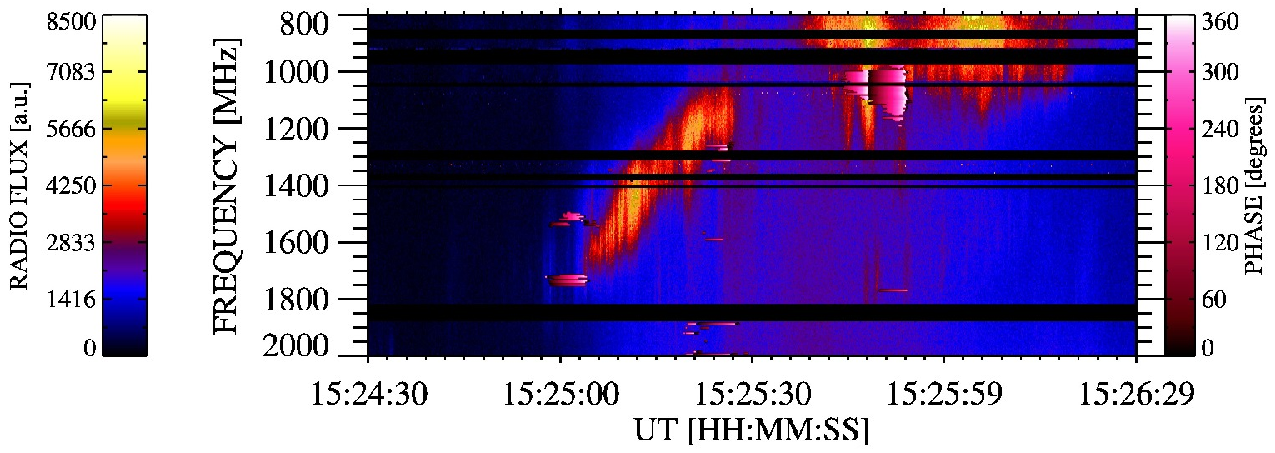}
\includegraphics[width=12.0cm,height=3.8cm, bb =30  10  400 135, clip=true]
%radio_base_WT_phase_overplot_freq_800-2000_MHz_time_152430-152630_per_7_2-10_5_s_lowres_arrows.eps}
{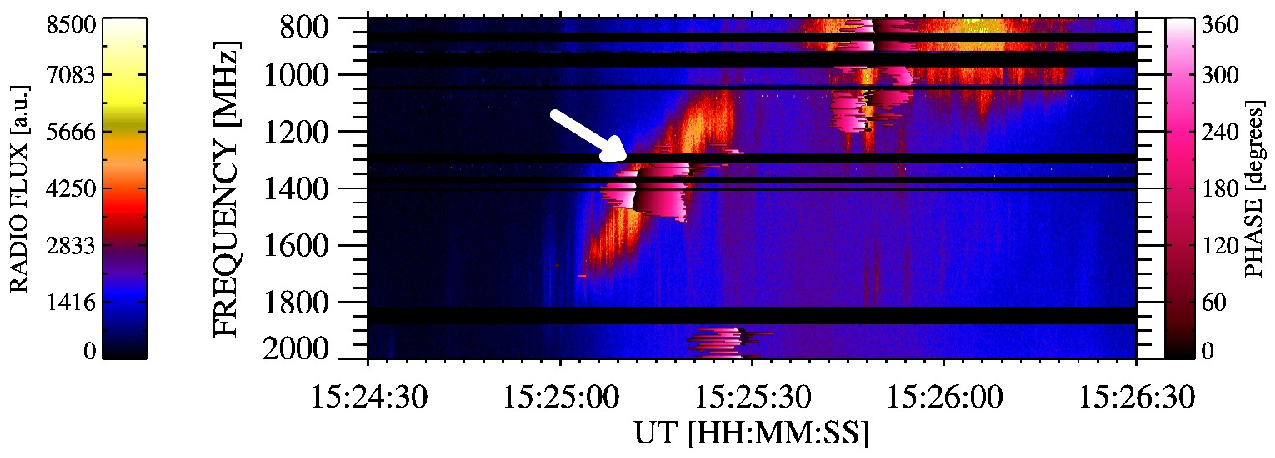}
\caption{DPS in the 20 May 2002 flare~--- continuation of Figure~\ref{figure3} for longer periods:
the phase maps (\textit{pink areas with the black lines} showing the zero phase of oscillations)
overplotted on the radio spectrum for periods detected in the range of 2.5--3.0, 3.1--4.3,
4.3--5.3, 5.4--6.2, and 7.2--10.5\,s (\textit{from up to down}).
The \textit{white arrow} shows selected distinct drifting oscillation phase.}
\label{figure4}
\end{figure}

\begin{figure}[t!]
\centering
% histograms
\includegraphics[width=12.0cm,height=4.0cm, clip=true]
%period_histogram_all_and_selected_freq_800-2000_MHz_time_124400-125400_per_1_0-12_0_s_lowres.eps}
{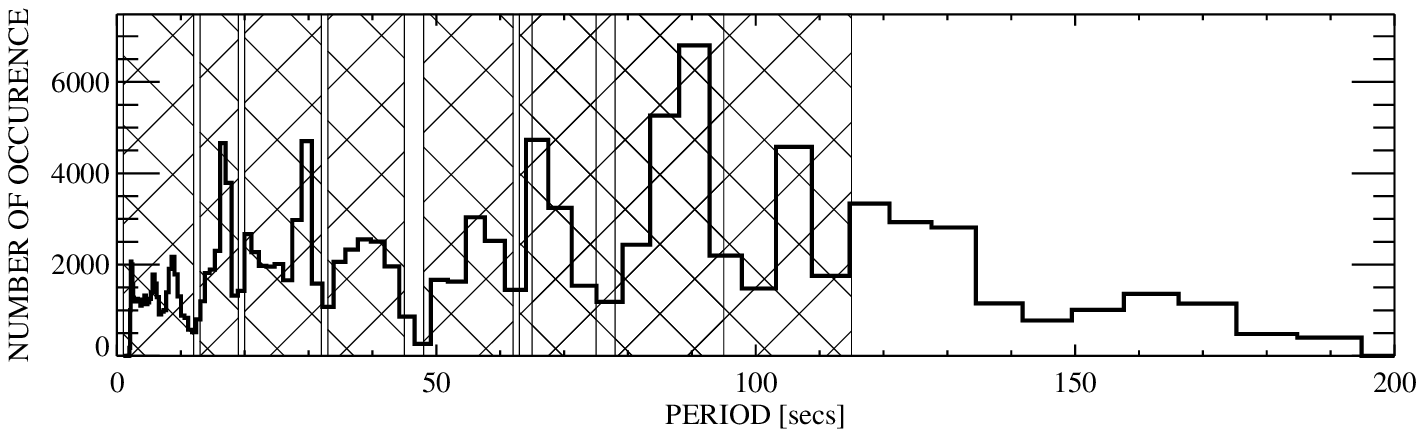}
% original spectrogram
\includegraphics[width=12.0cm,height=3.2cm, bb =30 30 420 135, clip=true]
%radio_base_WT_no_overplot_freq_800-2000_MHz_time_124400-125400_lowres.eps}
{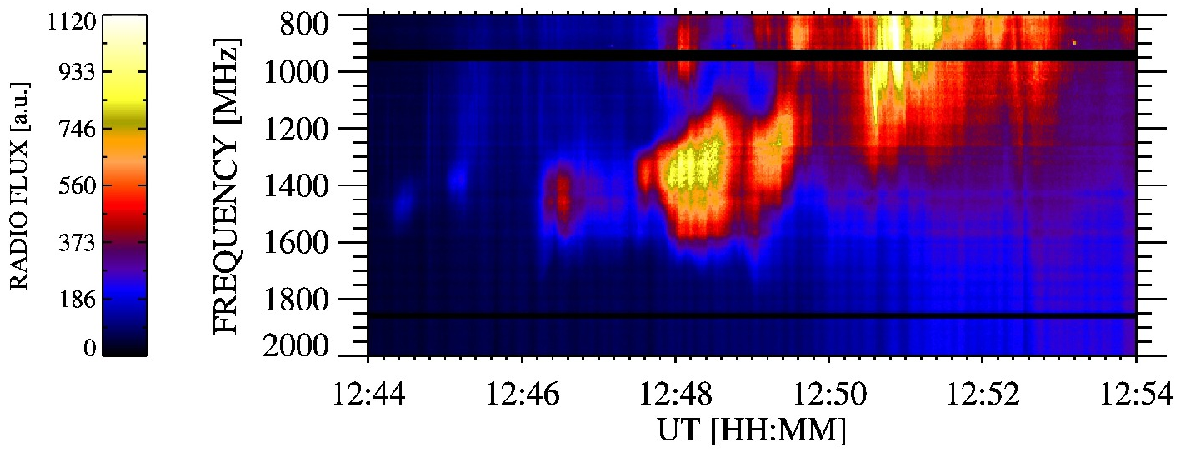}
% maps
\includegraphics[width=12.0cm,height=3.2cm, bb =30 30 420 135, clip=true]
%radio_base_WT_phase_overplot_freq_800-2000_MHz_time_124400-125400_per_1_0-12_0_s_lowres.eps}
{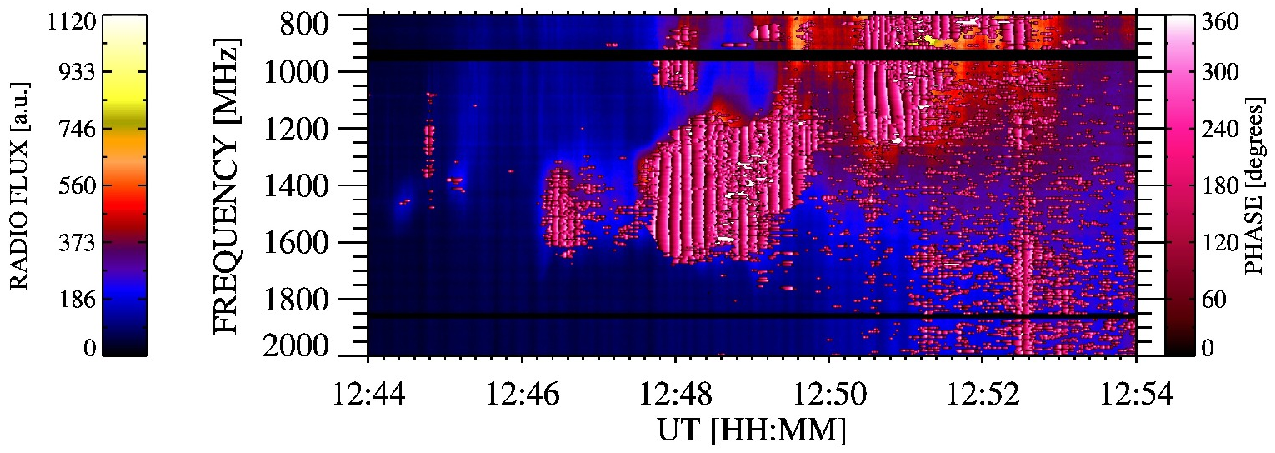}
\includegraphics[width=12.0cm,height=3.2cm, bb =30 30 420 135, clip=true]
%radio_base_WT_phase_overplot_freq_800-2000_MHz_time_124400-125400_per_13_0-19_0_s_lowres.eps}
{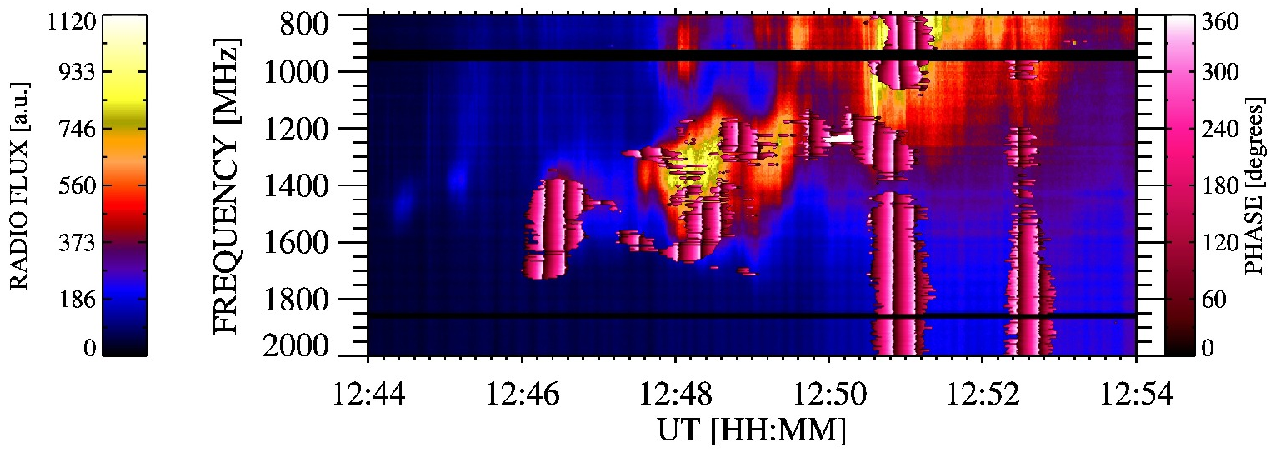}
\includegraphics[width=12.0cm,height=3.8cm, bb =30 10 420 135, clip=true]
%radio_base_WT_phase_overplot_freq_800-2000_MHz_time_124400-125400_per_20_0-32_0_s_lowres.eps}
{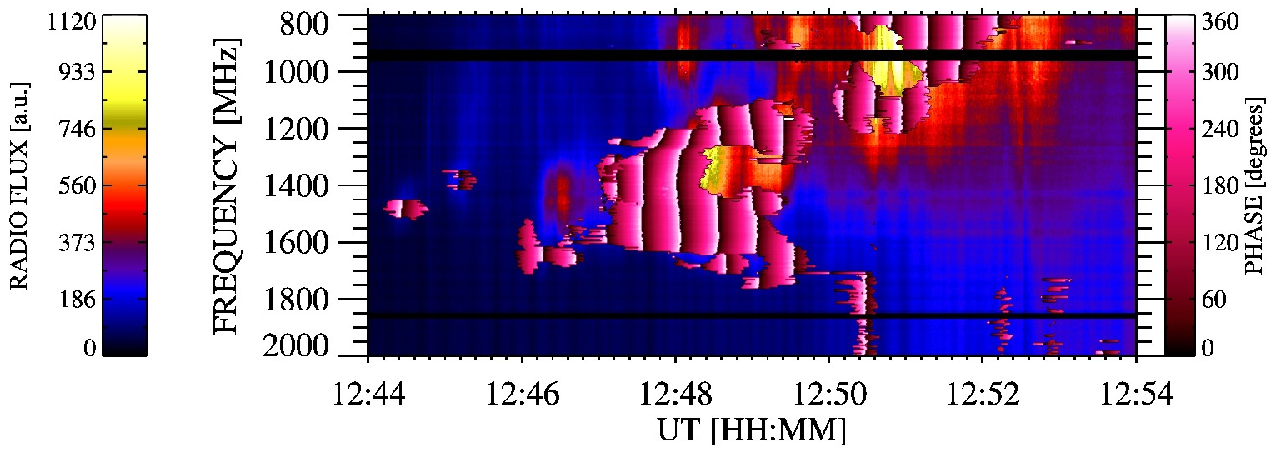}
\caption{DPS in the 18 April 2014 flare~--- time interval 12:44:00--12:54:00\,UT:
\textit{first panel} (\textit{from up to down}):
histogram of the WT significant periods, where the ranges of periods
selected for the following oscillation maps are expressed by \textit{hatched areas};
\textit{second panel}: the radio spectrum observed by
the \textit{Ond\v{r}ejov Radiospectrograph}; \textit{bottom three panels}: the
phase maps (\textit{pink areas with the black lines} showing the zero phase of oscillations)
overplotted on the radio
spectrum for periods in the range of 1--12, 13--19, and 20--32\,s.
} \label{figure5}
\end{figure}

\begin{figure}[t!]
\centering
\includegraphics[width=12.0cm,height=3.2cm, bb =30 30 420 135, clip=true]
%radio_base_WT_phase_overplot_freq_800-2000_MHz_time_124400-125400_per_33_0-45_0_s_lowres.eps}
{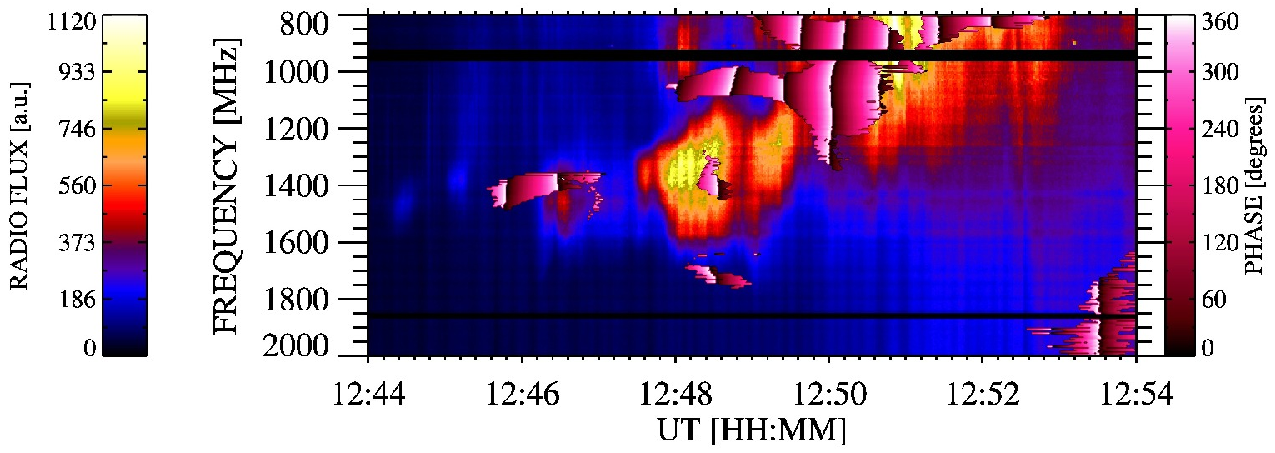}
\includegraphics[width=12.0cm,height=3.2cm, bb =30 30 420 135, clip=true]
%radio_base_WT_phase_overplot_freq_800-2000_MHz_time_124400-125400_per_48_0-62_0_s_lowres.eps}
{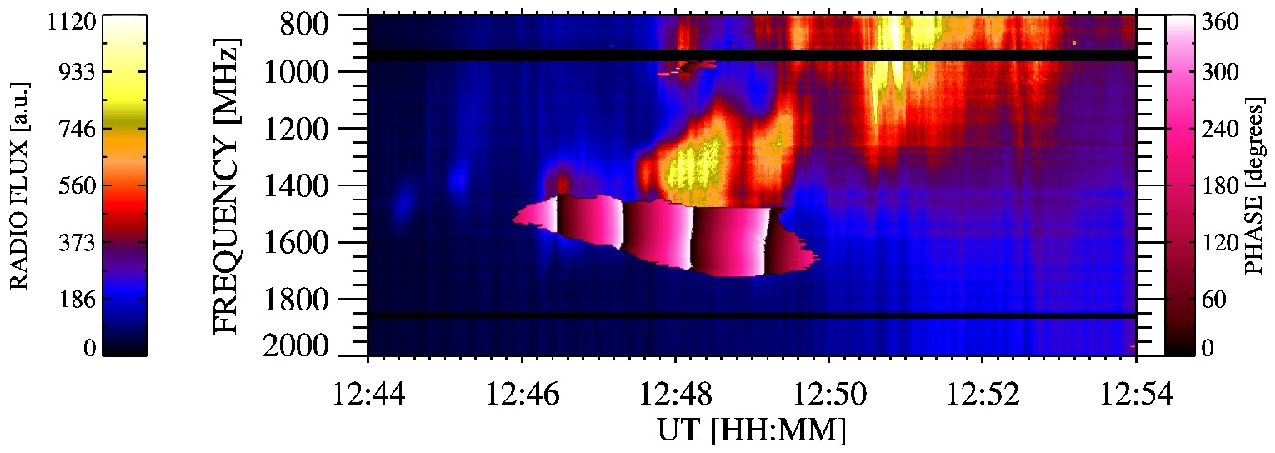}
\includegraphics[width=12.0cm,height=3.2cm, bb =30 30 420 135, clip=true]
%radio_base_WT_phase_overplot_freq_800-2000_MHz_time_124400-125400_per_63_0-75_0_s_lowres.eps}
{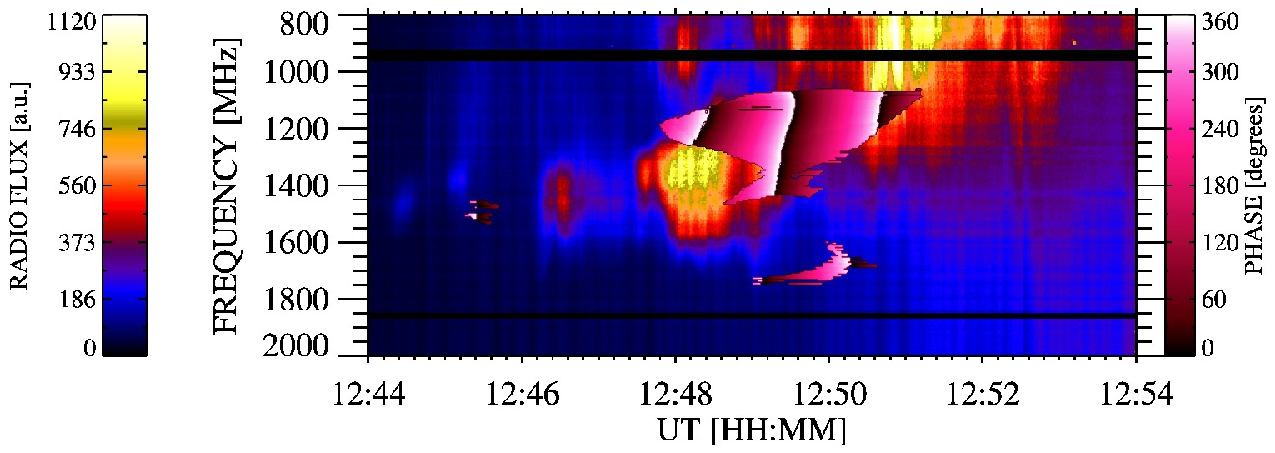}
\includegraphics[width=12.0cm,height=3.2cm, bb =30 30 420 135, clip=true]
%radio_base_WT_phase_overplot_freq_800-2000_MHz_time_124400-125400_per_78_0-95_0_s_lowres.eps}
{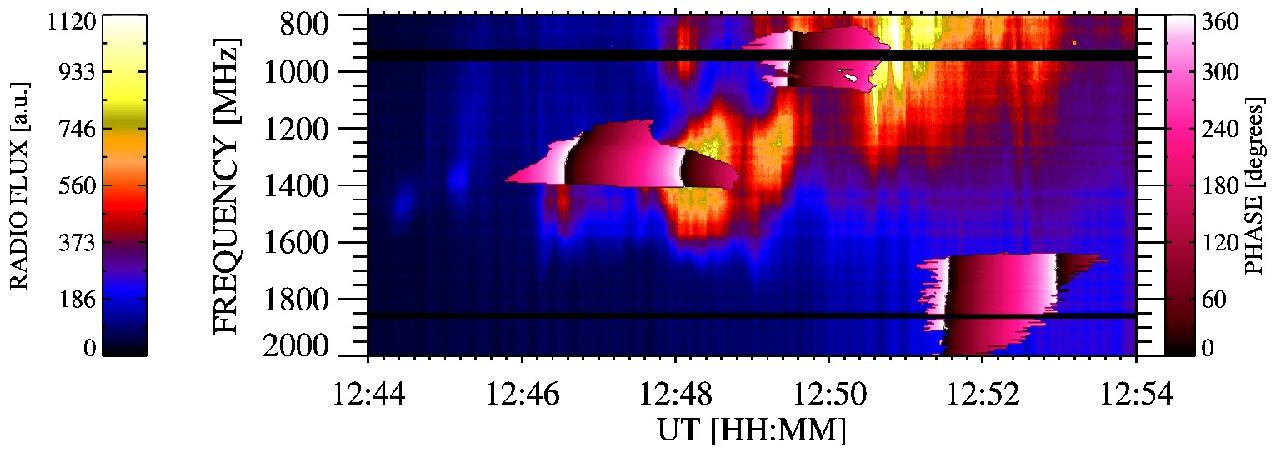}
\includegraphics[width=12.0cm,height=3.8cm, bb =30 10 420 135, clip=true]
%radio_base_WT_phase_overplot_freq_800-2000_MHz_time_124400-125400_per_65_0-115_0_s_lowres_arrows.eps}
{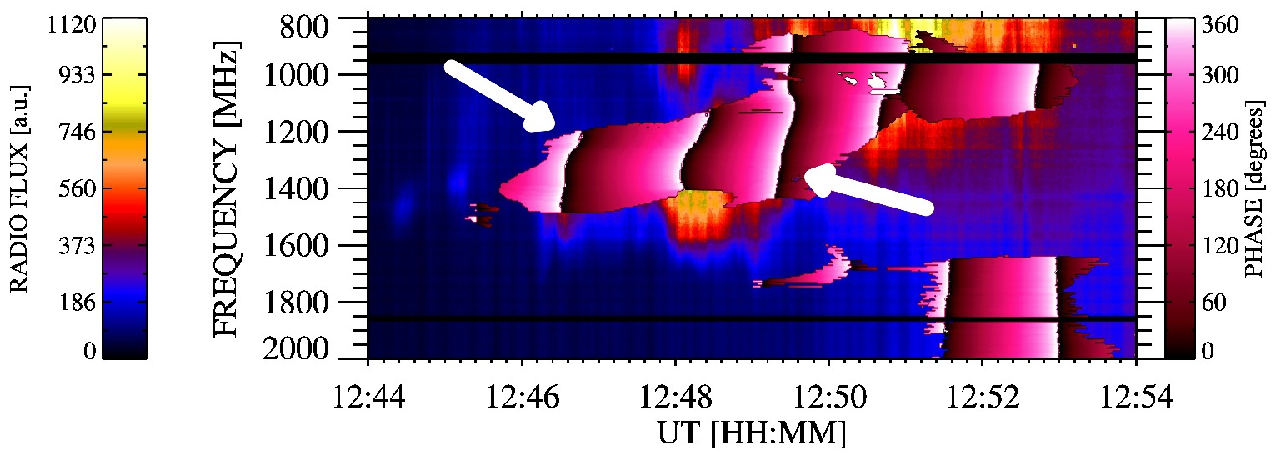}
\caption{DPS in the 18 April 2014 flare~--- continuation of Figure~\ref{figure5} for longer periods:
the phase maps (\textit{pink areas with the black lines} showing the zero phase of oscillations)
overplotted on the radio spectrum for periods detected in the range of 33--45, 48--62,
63--75, 78--95, and 65--115\,s (\textit{from up to down}).
Note that the last interval of periods (65--115\,s) covers the intervals of periods
with 63--75 and 78--95\,s.
The \textit{white arrows} show selected distinct drifting oscillation phases.}
\label{figure6}
\end{figure}

Examples of the DPS radio spectra together with the corresponding phase
oscillation maps are shown in Figures~\ref{figure1} and \ref{figure2} (DPS
No.1), \ref{figure3} and \ref{figure4} (DPS No.3), and \ref{figure5} and
\ref{figure6} (DPS No.7).

These figures start with the histogram of the WT significant periods
detected in the DPS at the top of the first figure. Peaks in these histograms
represent the main periods of oscillations found in the DPS which are
numerically listed in the second column of Table~\ref{tab3}. They were found
for periods in the 1--108\,s range. Around these peaks we selected the
ranges of periods for computation of the oscillation maps (see below the
discussion on how this period-range selection influences the results). The
next plot (from
up to down) shown in the figure is the DPS radiospectrum itself. The phase
oscillation maps, showing phase of the statistically significant oscillations
(pink areas with the black lines showing the zero phase) are shown then in
consecutive plots in these couples of figures starting from maps for the short
periods to those for longer periods.

In all DPSs, we found that at short periods the phases are more or less
synchronized; they have an "infinite" frequency drift. On the other hand, for
three DPSs (Nos.\,1, 3 and 7) and for longer periods, the phases of some
oscillations drift in frequencies. In Figures~\ref{figure1}--\ref{figure6}
the most distinct drifting phases are designated by white arrows and their
parameters are summarized in Table~\ref{tab3}. Drifts of these oscillation
phases (DPHA) were found to be in the range from $-17$ to $+287$MHz\,s$^{-1}$.

To show how these oscillation maps depend on the selected interval of periods,
in Figure~\ref{figure6} we present not only the phase maps for distinct period
intervals of the 63--75 and 78--95\,s (third and fourth panels),
but also the map for the 65--115\,s period (the bottom panel), which
covers both these 63--75 and 78--95\,s period ranges. Comparing
these maps we can see how phases presented in the map with a broader range of
periods are distributed in the time--frequency domain in the maps with narrower
ranges of periods.

Furthermore, comparing the parameters of DPSs (Table~\ref{tab2}) with those of
oscillation periods and drifting phases DPHA we can see that the DPS No.\,7
(18 April 2014) differs from other DPSs. This DPS has not only the smallest
frequency drift ($-1.66$\,MHz\,s$^{-1}$), but also much longer main periods (up
to 108\,s) as well as smaller absolute frequency drift of drifting phases
($-17$ and $-21$\,MHz\,s$^{-1}$) comparing to other DPSs. In other DPSs the
interval of oscillation periods is up to 15\,s, approximately.

%-------------------------------------------------------------------------------
% 19921005 - 2nd time interval 09:24:10-09:25:00

%----------------------------------------------------------------------------
% 20020520 - time interval: 15:24:30,15:26:30

%----------------------------------------------------------------------------
% 2014/04/18 - time interval: 12:40 - 13:10 cadence 1s

%%%%%%%%%%%%%%%%%%%%%%%%%%%%%%%%%%%%%%%%%%%%%%%%%%%%%%%%%%%%%%%%%%%%%%%%%%%%%%%%
\section{Interpretation}

In this work we are interested not only about periods of oscillations in
DPSs, but also about the frequency drifts of phases of these oscillations.
Namely, DPSs are generated by the plasma emission mechanism
\citep{2008SoPh..253..173B}, therefore the emission frequency is connected with
the plasma density and thus with the altitude  of the radio source in the
gravitationally stratified solar atmosphere. Therefore, the very high frequency
drift of oscillation phases can be interpreted by the high speed of the
superthermal electrons injected into the plasmoid. On the other hand, their low
frequency drift can indicate a presence of some magnetohydrodynamic waves or
plasma density structures in or close to the plasmoid.

Analyzing seven DPSs by a new type of oscillation maps, we found
oscillations with periods in the range 0.5--13.3\,s for six DPSs (Nos.\,1--6)
and in the range 7--108\,s for DPS No.\,7. The frequency drift of the DPSs
No.\,1--6 ( from $-5.5$ to $-30$\,MHz\,s$^{-1}$) was greater than
that of the DPS No.\,7 ($-1.66$\,MHz\,s$^{-1}$).

We found that oscillations especially for short periods are
practically synchronized ("infinite" frequency drift). This
synchronization with the period of about 1\,s confirms findings in our previous
studies \citep{2000A&A...360..715K,2014RAA....14..753K} and shows that the used
wavelet method works correctly. In agreement with our previous papers we
interpret this synchronization as caused by a quasi-periodic injection of fast
superthermal electrons into the plasmoid. A broad range of periods of the
oscillations with "infinite" drift is a new result and we think that it is
caused by fragmented magnetic reconnection, where particles are
accelerated during the merging of plasmoids of different sizes and thus with
different periods \citep{2011ApJ...733..107K,2011ApJ...737...24B}.

In DPSs No.\,1, 3 and 7 for longer oscillation periods we found the
oscillations with drifting phases. In DPS No.\,1 the frequency drift of
drifting phase (oscillation period $6.5$--$8.5$\,s) was positive ($+287$ and
$+125$\,MHz\,s$^{-1}$) , in DPS No.\,3 (oscillation period 7.2--10.5\,s)
negative ($-125$\,MHz\,s$^{-1}$), and in DPS No.\,7 (oscillation period
65--115\,s) negative ($-17$ and $-21$ MHz\,s$^{-1}$), but its absolute value
was much smaller than in the DPSs No.\,1 and 3.

As shown in the papers by \cite{2007CEAB...31..165B} and
\cite{2017ApJ...847...98J} magnetic reconnection is closely connected with
magnetosonic waves, which are generated at regions of enhanced
(anomalous) resistivity appear. At these regions the plasma is strongly heated,
its pressure increases and thus the magnetosonic waves are generated and
propagate out of these regions. Generally, there are many such regions around
X-points of the fragmented magnetic reconnection, therefore it is highly
probable that some of these will interact with the plasmoid producing the DPS.
Such waves can modulate the plasma density and thus modulate a DPS emission, as
described {\bf {\it e.g.}} in the papers by \cite{2013A&A...550A...1K} and
\cite{2013A&A...552A..90K}.

Based on these results, the frequency drift of the drifting oscillation phase can
be interpreted as caused by magnetosonic waves in (or in close vicinity of) the
plasmoid. If we take the size of the plasmoid (where DPS is generated) as in the
papers by \cite{1998ApJ...499..934O, 2010SoPh..266...71K} (10 or 25~Mm) and the
Alfv\'en speed at the plasmoid as $1000$\,km\,s$^{-1}$, approximately, then the drifting
oscillation phases in DPSs No.\,1 and 3 can be caused by the fast magnetosonic
wave. Their positive and negative frequency drifts are then caused by the waves
propagating through the plasmoid in the direction of the density gradient or in
the opposite direction.

Period of drifting oscillation phases in DPS No.\,7 are too long to be
interpreted as in the case of DPSs No.\,1 and 3. We propose that in this DPS No.\,7
the drifting oscillation phases are caused by the slow magnetosonic waves.

Besides interpretation of the drifting phases by the magnetosonic waves, there
are also other possible interpretations, {\bf {\it e.g.}} by a quasi-periodic structure in
the plasma inflowing to the reconnection forming the plasmoid and by a
quasi-periodically varying reconnection rate in the X-point of the reconnection
close to the plasmoid. We think that both these processes generate the
quasi-periodic density structure of the plasmoid and this structure, as the
plasmoid grows, can modulate the DPS.

\section{Conclusions}

This study gives us new and more detailed information about oscillations and
waves in and around the plasmoid formed during the flare magnetic reconnection.

We analyzed seven DPSs observed by the \textit{Ond\v{r}ejov Radiospectrograph}. 
Using a new type of oscillation maps, we found that DPSs are very rich in
oscillations in the range of 1--108\,s periods. This strongly supports the
idea that DPSs are generated in the plasmoid during fragmented magnetic
reconnection.

In all DPSs, especially for shorter periods, the oscillations were synchronized
in the whole range of DPSs. In agreement with a theoretical interpretation of
the DPSs, we propose that these synchronized oscillations are caused by
quasi-periodic injections of the superthermal electrons into the plasmoid from
fragmented magnetic reconnection.

On the other hand, in some DPSs and mostly for longer periods we found
oscillations having their phases drifting in frequencies. In assumed plasma
emission mechanism of DPSs it indicates a presence of some magnetosonic waves
in and around the plasmoid or some evolving density structure in the plasmoid.
We propose that the oscillations with drifting phases can be caused (a) by
fast or slow magnetosonic waves, generated during fragmented magnetic
reconnection and propagating through the plasmoid, (b) by a quasi-periodic
structure in the plasma inflowing to the reconnection region forming the plasmoid, 
and (c) by a quasi-periodically varying reconnection rate in the X-point of
reconnection close to the plasmoid.

%----------------------------------------------------------------------------------------------------
\begin{acks}
% CZ
This research was supported by Grants 16-13277S and 17-16447S of the Grant
Agency of the Czech Republic.
% SK
This work was supported by the Science Grant Agency project VEGA 2/0004/16 (Slovakia).
% MAD
Help of the Bilateral Mobility Projects SAV-16-03 and SAV-18-01
of the SAS and CAS is acknowledged.
%SF EU
This article was created in the project ITMS No. 26220120029, based on the
supporting operational Research and development program financed from the
European Regional Development Fund.
% general:
% ADS/NASA
This research has made use of NASA's Astrophysics Data System.
% T+C
The wavelet analysis was performed with software based on tools provided by C.
Torrence and G. P. Compo at {\tt http://paos.colorado.edu/research/wavelets}.
\end{acks}

\section*{Disclosure of Potential Conflicts of Interest}
The authors declare that they have no conflicts of interest.

%%----------------------------------------------------------------------------------------------------
%\bibliographystyle{spr-mp-sola}
%\bibliography{karlicky_rybak_barta_dps}

%----------------------------------------------------------------------------------------------------
\end{article}
\end{document}